\documentclass[prl,twocolumn,showpacs,preprintnumbers,amsmath,amssymb]{revtex4}

\usepackage{graphicx}
\usepackage{dcolumn}
\usepackage{bm}

\def\lsim{\mathrel{\lower2.5pt\vbox{\lineskip=0pt\baselineskip=0pt
\hbox{$<$}\hbox{$\sim$}}}}
\def\gsim{\mathrel{\lower2.5pt\vbox{\lineskip=0pt\baselineskip=0pt
\hbox{$>$}\hbox{$\sim$}}}}

\newcommand{\be}{\begin{equation}}
\newcommand{\ee}{\end{equation}}

\begin{document}


\title{Precise determination of the $f_0(600)$ and $f_0(980)$ pole
parameters from a dispersive data analysis}

\author{R. Garc\'{\i}a-Mart\'{\i}n$^a$, R.~Kami\'nski$^b$, J. R. Pel\'aez$^a$ and J. Ruiz de Elvira$^a$}
\affiliation{
$^a$Departamento de F{\'\i}sica Te{\'o}rica II,
 Universidad Complutense de Madrid, 28040   Madrid,\ Spain\\
$^b$Department of Theoretical Physics
Henryk Niewodnicza\'nski Institute of Nuclear Physics,
Polish Academy of Sciences,
31-342, 
Krak\'ow, Poland.}


\begin{abstract}
We use our latest dispersive analysis of $\pi\pi$ scattering data and 
the very recent $K_{\ell 4}$ experimental results to obtain the mass,
width and couplings of the two lightest 
scalar-isoscalar resonances. These parameters are defined from 
their associated poles in the complex plane. The analytic continuation
to the complex plane is made in a model independent way by means of
once and twice subtracted dispersion relations for the partial waves,
without any other theoretical assumption.
We find the $f_0(600)$ pole at  $(457^{+14}_{-13})-i(279^{+11}_{-7})$~MeV
and that of the $f_0(980)$ at $(996\pm7)-i(25^{+10}_{-6})$ MeV,
whereas their respective couplings to two pions are $3.59^{+0.11}_{-0.13}$ GeV and 
$2.3\pm0.2$ GeV.
\end{abstract}

\maketitle


The $f_0(600)$ or sigma and $f_0(980)$ resonances
are of great interest in several fields of Physics.
First,  the
two pion exchange in the scalar-isoscalar channel, I=0, J=0, 
where these resonances appear, plays a key role in Nuclear Physics,
where the nucleon-nucleon
attractive interaction has been for long \cite{Johnson:1955zz}
modeled by the exchange of a ``sigma'' resonance. 
Second, this channel is also relevant for the QCD non-Abelian nature,
since it is where the lightest glueball
is expected to appear. However,  
the glueball identification is complicated by its possible mixing
into different states,
like the $f_0(600)$, $f_0(980)$ and heavier $f_0$ resonances, which may be
$\bar q q$ mesons, 
tetraquarks, molecules, or most likely a mixture of them all.
Actually, most of the controversy around these resonances
comes from the identification of scalar multiplets
---see the Review of Particle Physics (PDG) 
``Note on Scalar Mesons''\cite{PDG}. Third, the $f_0(600)$, being the lightest 
hadronic resonance with vacuum quantum
numbers, plays 
a relevant role in many models of QCD spontaneous chiral 
symmetry breaking. 
Furthermore,  this state is of interest 
in order to understand why, despite being so light and strongly
coupled to pions, it plays such a small
role, if any, in the saturation \cite{Ecker:1988te} of the 
low energy constants of Chiral Perturbation Theory (ChPT). 
Moreover, the position of this pole could
be setting the limit of applicability of the chiral expansion. 
Finally, this state is of interest for Electroweak Physics
due to its many similarities 
---but even more by its many differences---with the Higgs mechanism 
now under scrutiny at LHC.

Still, the properties of these resonances
are the subject of an intense debate. Let us recall that the $\sigma$ was listed in the PDG as
``not well established'' until 1974, removed in 1976, 
and listed back in 1996.
This was due to its width being
comparable to its mass, so that it barely propagates and  
becomes a broad enhancement in the traditional, and often contradictory,
 $\pi\pi$ scattering analyses, extracted from 
$\pi N\rightarrow \pi\pi N$ experiments, using different models 
affected by large systematic uncertainties. 
After 2000 these resonances have been observed in
decays of heavier mesons, with well defined initial states 
and very different systematics
from $\pi\pi$ scattering, which led the PDG to consider, in 2002, 
the $f_0(600)$  as ``well established'',
but keeping until today a too conservative
estimate of:
``Mass: 400 to 1200 MeV'' and ``Width: 600 to 1000 MeV''.
For the $f_0(980)$ the situation is not much better, with an estimated width
 ``from 40 to 100 MeV''. However, not all the uncertainty comes from experiment.
The shape of these resonances varies from process to process
and that is why their masses and widths are 
quoted from their process independent pole
positions, defined as
$\sqrt{s_{\rm pole}}\sim M- i\Gamma/2$. 
But many models do not 
implement rigorous analytic continuations
and lead to incorrect determinations when poles are deep in the complex plane or close to 
threshold cuts, as it happens with the $f_0(600)$ and the $f_0(980)$, respectively.
Actually, this is one of the main causes of the huge PDG uncertainties
\cite{PDG}.

This model dependence can be avoided by using dispersive techniques, which follow from
causality and crossing, and
provide integral relations
 and a rigorous analytic continuation of the
amplitude in terms of  its imaginary
part in the physical region, which can be obtained from data.
For example, dispersion relations combined
with ChPT determine the $\sigma$ pole at
$440 - i\,245~\mbox{MeV}$ \cite{Dobado:1996ps} or 
$(470\pm50) - i\,(260\pm25)~\mbox{MeV}$ \cite{Zhou:2004ms}. 
We focus here on dispersive analyses,
but  other approaches yield similar values
\cite{sigma,Oller:1998hw}---see Table \ref{tab:sigmapolecoupling}
and \cite{Klempt:2007cp} for a review and references.

\begin{table}
  \centering
  \begin{tabular}{|c|c|c|}
\hline
 & $\sqrt{s_\sigma}$ (MeV) & $\vert g_{\sigma\pi\pi}\vert$ (GeV) \\ \hline
\rule[-1mm]{0mm}{4mm}\cite{Caprini:2005zr}& $441^{+16}_{-8}-i (272^{+9}_{-12.5})$&$3.31^{+0.35}_{-0.15}$\\ \hline
\cite{Yndurain:2007qm}& $474\pm6-i (254\pm4)$& $3.58\pm0.03$\\ \hline
\rule[-1.5mm]{0mm}{5mm}\cite{Caprini:2008fc}& $463\pm6^{+31}_{-17}-i (254\pm6^{+33}_{-34})$& -\\ \hline
\cite{Oller:2003vf}& $(443\pm2)-i(216\pm4)$ & $2.97\pm0.04 $\\\hline
\cite{Mennessier:2010xz}& $452\pm12-i 260\pm15$& $2.65\pm0.10$\\ \hline 
\cite{Pelaez:2010fj} (fit D) & $453-i\,271$ & 3.5 \\
\hline
  \end{tabular}
  \caption{Other recent determinations of the $\sigma$ pole and 
    coupling, using analyticity. Results come from Roy eqs. and ChPT \cite{Caprini:2005zr}, 
 conformal fits to $K_{\ell4}$ decays and
averaged $\pi\pi$ data around 800-900 MeV with only statistical \cite{Yndurain:2007qm} 
or also systematic \cite{Caprini:2008fc} uncertainties, 
the chiral unitary approach \cite{Oller:2003vf}
(only statistical error), a $K$-matrix with
a form factor shape \cite{Mennessier:2010xz}, and
ChPT+elastic dispersion relations 
(two-loops \cite{Pelaez:2010fj}). }
\label{tab:sigmapolecoupling}
\end{table}

\begin{table}
  \centering
  \begin{tabular}{|c|c|c|}
\hline
 & $\sqrt{s_{f_0(980)}}$ (MeV) & $\vert g_{f_0\pi\pi}\vert$ (GeV) \\ \hline
\cite{Akhmetshin:1999di} & $(978\pm12)-i(28\pm15)$&$2.25 \pm 0.20$\\\hline
\cite{Barberis:1999cq}& $(988\pm10\pm6)-i(27\pm6\pm5)$& $2.2\pm0.2$ \\\hline
\cite{Aitala:2000xt}  & $(977\pm5)-i(22\pm2)$& $1.5\pm0.2$\\\hline
\cite{Ablikim:2004wn}& $(965\pm10)-i(26\pm11)$& $2.3\pm0.2$\\\hline
\cite{Oller:2003vf} & $(986\pm 3)-i(11\pm 4)$& $1.1\pm0.2$\\\hline
\cite{Mennessier:2010xz}& $(981\pm34)-i(18\pm11)$&$1.17\pm0.26$\\\hline
\cite{Mao:2009cc}& $999-i\,21$&  1.88\\\hline
  \end{tabular}
  \caption{Recent determinations of $f_0(980)$ parameters.
For \cite{Barberis:1999cq} our estimate covers the six models 
considered there. The last three poles come from scattering matrices
and the rest from production experiments.}
\label{tab:f0polecoupling}
\end{table}

Generically, the main difficulty lies in the calculation of the left cut
integral, which in \cite{Dobado:1996ps,Zhou:2004ms} was just approximated.
This left cut is due to crossing symmetry
and can be incorporated rigorously in a set of infinite coupled
equations written long ago by Roy~\cite{Roy:1971tc} ( see also
\cite{roy70} for applications and references). 
Recently, Roy equations have been used
to study low energy 
$\pi\pi$ scattering 
\cite{Ananthanarayan:2000ht}, sometimes combined with ChPT \cite{Colangelo:2001df},
or also to test ChPT ~\cite{DescotesGenon:2001tn}, as well as to
solve old data ambiguities \cite{Kaminski:2002pe}.
Most recently \cite{Caprini:2005zr}, the $f_0(600)$ and $f_0(980)$ poles 
were shown to lie within the applicability region of Roy eqs. 
Since data were not reliable and to improve accuracy, Roy eqs.
were supplemented by ChPT predictions in \cite{Caprini:2005zr}, to yield :
$  \sqrt{s_\sigma}= 441_{-8}^{+16}-i\,272_{-19.5}^{+9}\mbox{ MeV}$,
without using data below 800 MeV on S and P waves.
In that work an $f_0(980)$ pole is also found at $\sqrt{s}= 1001-i\,14\,\mbox{ MeV}$.
Note that, generically, $\pi\pi$ scattering data around 900 MeV tend to produce a
narrower $f_0(980)$ \cite{Oller:2003vf,Oller:1998hw,Caprini:2005zr} 
than that seen in 
production processes or the PDG estimate. In Table~\ref{tab:f0polecoupling} we list
some other recent determinations of the $f_0(980)$ parameters.

Our aim in this work is to provide a precise and model independent simultaneous
determination of the $f_0(600)$ and $f_0(980)$ parameters from data alone, profiting
from two relevant results developed over the 
last half year: On the one hand, 
the  final analysis of $K_{\ell 4}$ decays by the
NA48/2 Collaboration \cite{ULTIMONA48}, which provides reliable and precise $\pi\pi$ scattering
phases below the mass of the kaon. On the other hand, 
a set of Roy-like eqs.---called GKPY eqs. and developed by our group \cite{GarciaMartin:2011cn}--- 
which is much more
stringent in the resonant region than standard Roy eqs.
The reason is that, in order to avoid 
divergences, dispersion relations are weighted at low energy 
with ``subtractions'', but then amplitudes are only determined up to a polynomial,
whose coefficients depend on threshold parameters.
Since Roy eqs. have two subtractions they have an $s$ polynomial term
multiplied by the isospin 2 scalar scattering length, whose large uncertainty 
thus grows markedly in the $f_0(600)$ and $f_0(980)$ region.
In contrast, the GKPY eqs. have just one subtraction and their output, even without using
ChPT predictions at all, provides \cite{GarciaMartin:2011cn} a very precise 
description of $\pi\pi$ scattering data, discarding a long-standing conflict
concerning the inelasticity---and to a lesser extent the phase shift---right above the $f_0(980)$ region.

If we now use these GKPY dispersion relations to continue analytically
that amplitude, we find:
\begin{eqnarray}
\sqrt{s_\sigma}&=&(457^{+14}_{-13})-i(279^{+11}_{-7})\; {\rm MeV}\\
\sqrt{s_{f_0(980)}}&=&(996\pm7)-i(25^{+10}_{-6})\; {\rm MeV}.
\end{eqnarray}
Let us describe next the whole approach in detail and provide
determinations for other quantities of interest, like their couplings and
the $\rho(770)$ parameters, as well as other checks of our calculations from Roy eqs.

\begin{figure}
\includegraphics[scale=.23]{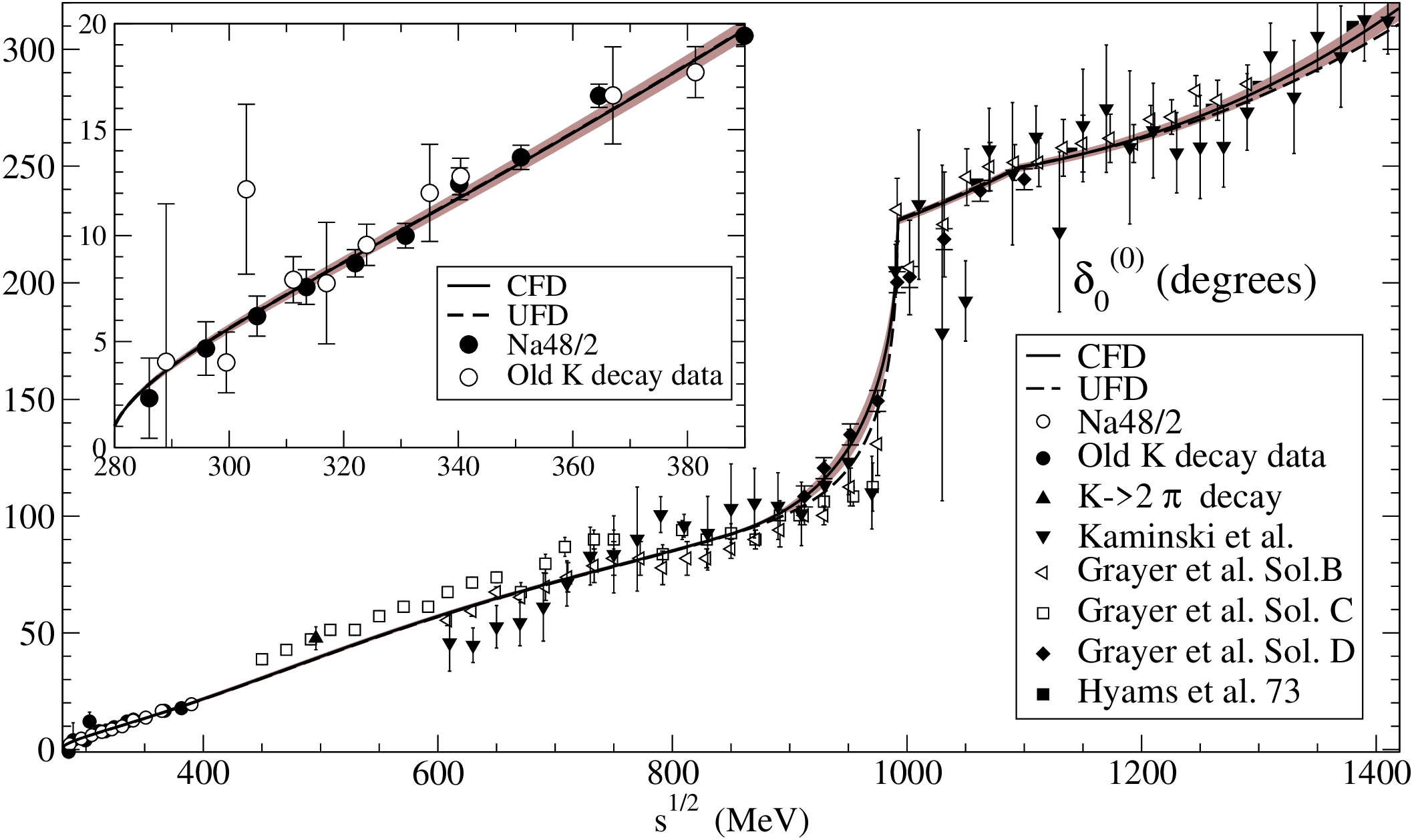}
\\
\includegraphics[scale=.23]{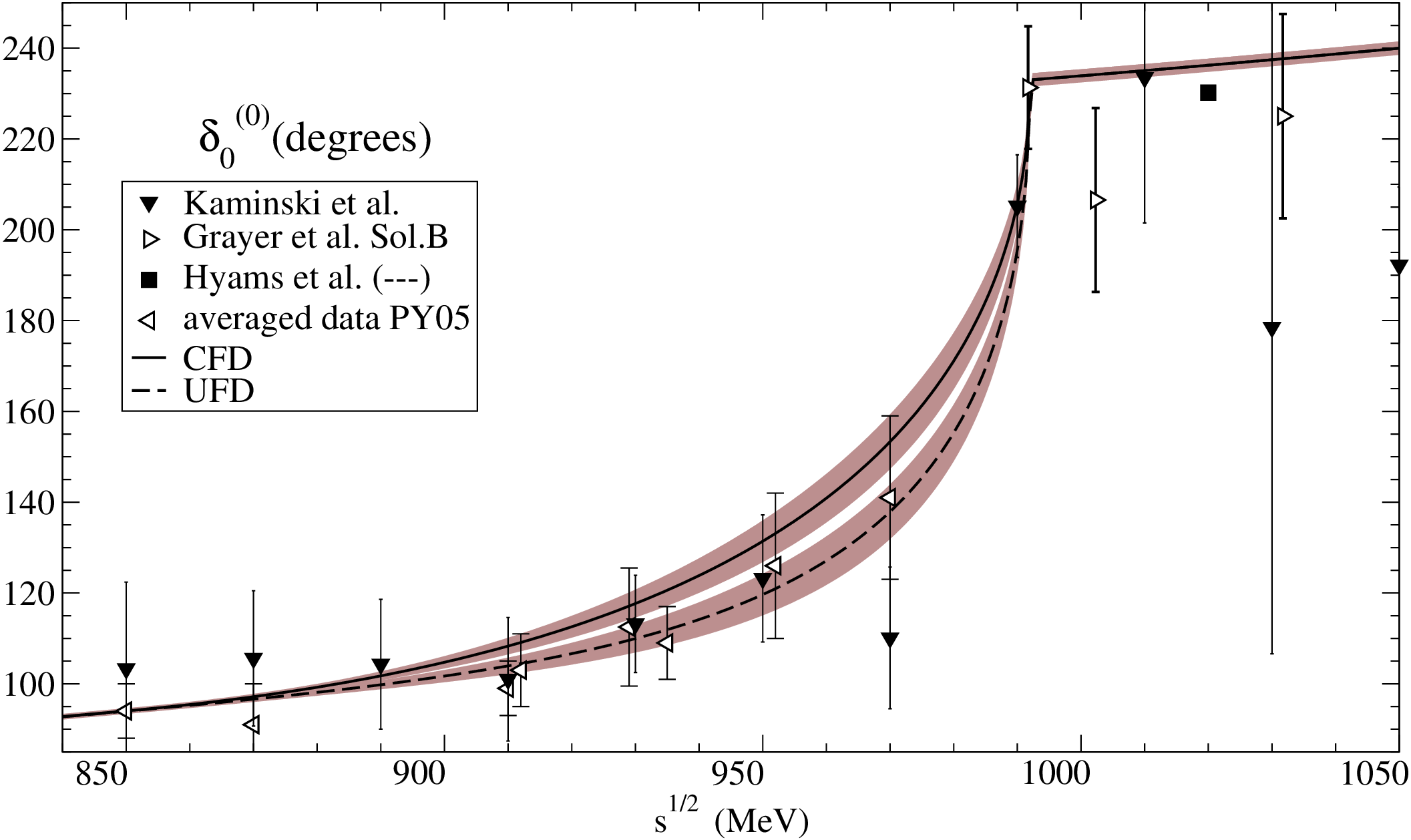}
\\
\includegraphics[scale=.23]{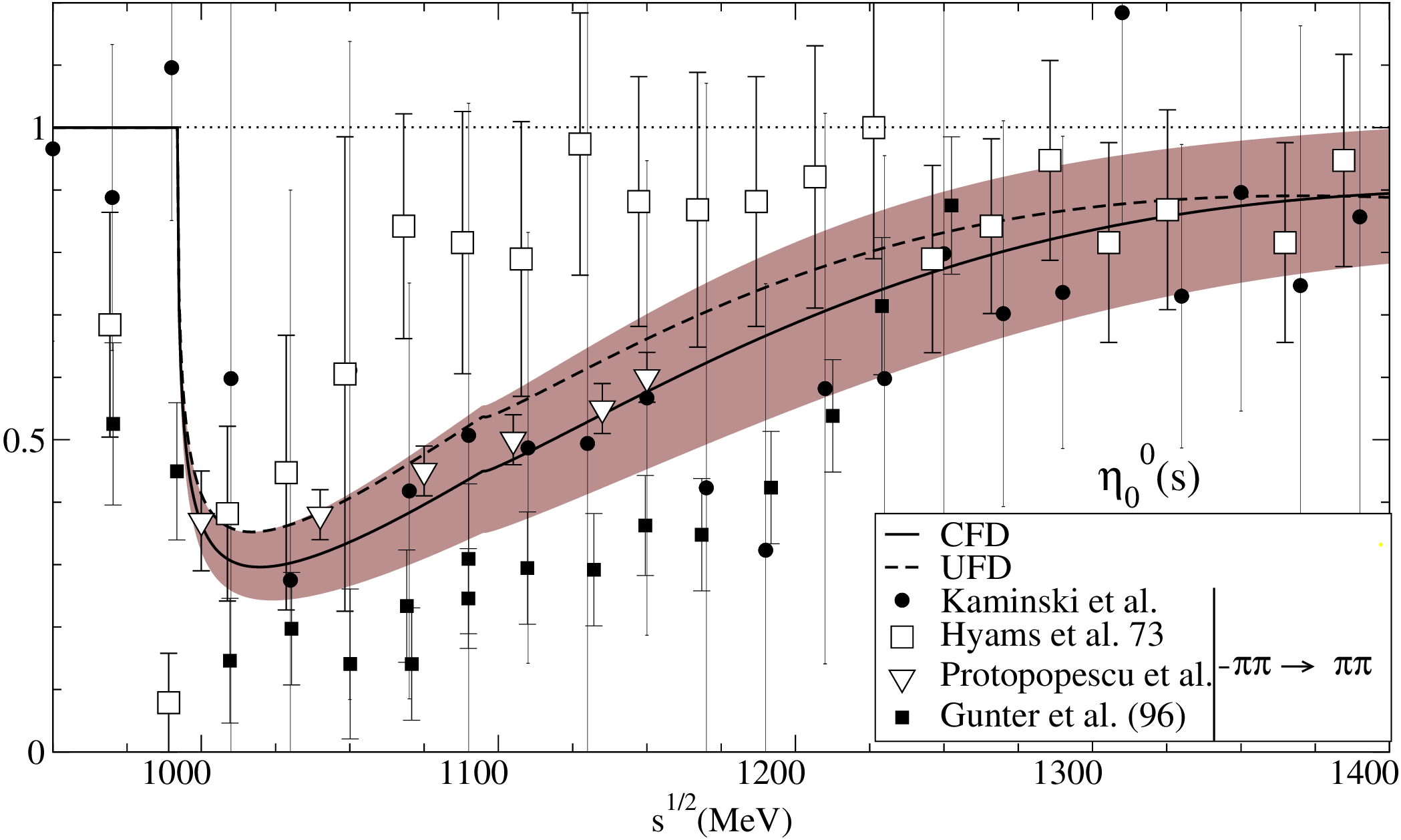}
\caption{\rm S0 wave phase and inelasticity from UFD and CFD.
Dark bands cover the uncertainties. Data come from \cite{ULTIMONA48,pipidata}.}
\label{S0wave}
\end{figure}

Ours is
what is traditionally called an
``energy-dependent''  analysis of
$\pi\pi$ scattering and $K_{\ell 4}$ decay data \cite{pipidata,Rosselet:1976pu}---in particular the latest results from NA48/2~\cite{ULTIMONA48}. 
Our procedure, described in a series of works \cite{Kaminski:2006qe,GarciaMartin:2011cn} was first to obtain
a simple set of {\it unconstrained} fits to these data (UFD) for each partial wave separately up to 1420 MeV, and Regge fits above that energy.
Next we obtained {\it constrained} fits to data (CFD) by varying the UFD parameters in order to satisfy within uncertainties two crossing sum rules,  a complete set of Forward Dispersion Relations
as well as Roy and GKPY eqs., while simultaneously describing data. 
The details for all CFD waves can be found in \cite{GarciaMartin:2011cn}, but since we are now interested in the scalar isoscalar partial wave $t_0^{(0)}$, we show in Fig.~\ref{S0wave} the resulting $\delta_0^{(0)}$ phase shift.
It should be noticed that the CFD result is indistinguishable to the eye from the UFD, except in the 900 to 1000 MeV region, which we also show in detail and 
is essential for the determination of the $f_0(980)$ parameters.  
Note that both the UFD and CFD describe the 
data in that region, but the 
GKPY dispersion relations require the CFD phase to lie somewhat higher 
than the UFD one. This is relevant since it yields a wider $f_0(980)$, correcting the above mentioned tendency to obtain
a too narrow $f_0(980)$ from unconstrained fits to $\pi\pi$ scattering data alone. 
In the inner top panel, we  show the good description of the 
latest NA48/2 data on $K_{\ell 4}$
decays, which are responsible for the small 
uncertainties in our input parametrization and constrain our
subtraction constants.
As seen in Fig.~\ref{S0wave}, the inelasticity $\eta_0^{(0)}$ shows
a ``dip'' structure 
above 1~GeV required by the GKPY eqs. \cite{GarciaMartin:2011cn}, which disfavors
the alternative ``non-dip'' solution. Having this long-standing ``dip'' versus ``no-dip''
controversy \cite{Au:1986vs} settled \cite{GarciaMartin:2011cn}
 is very relevant for a precise $f_0(980)$ determination.

\begin{figure}
\hspace*{-.4cm}
\includegraphics[scale=.32]{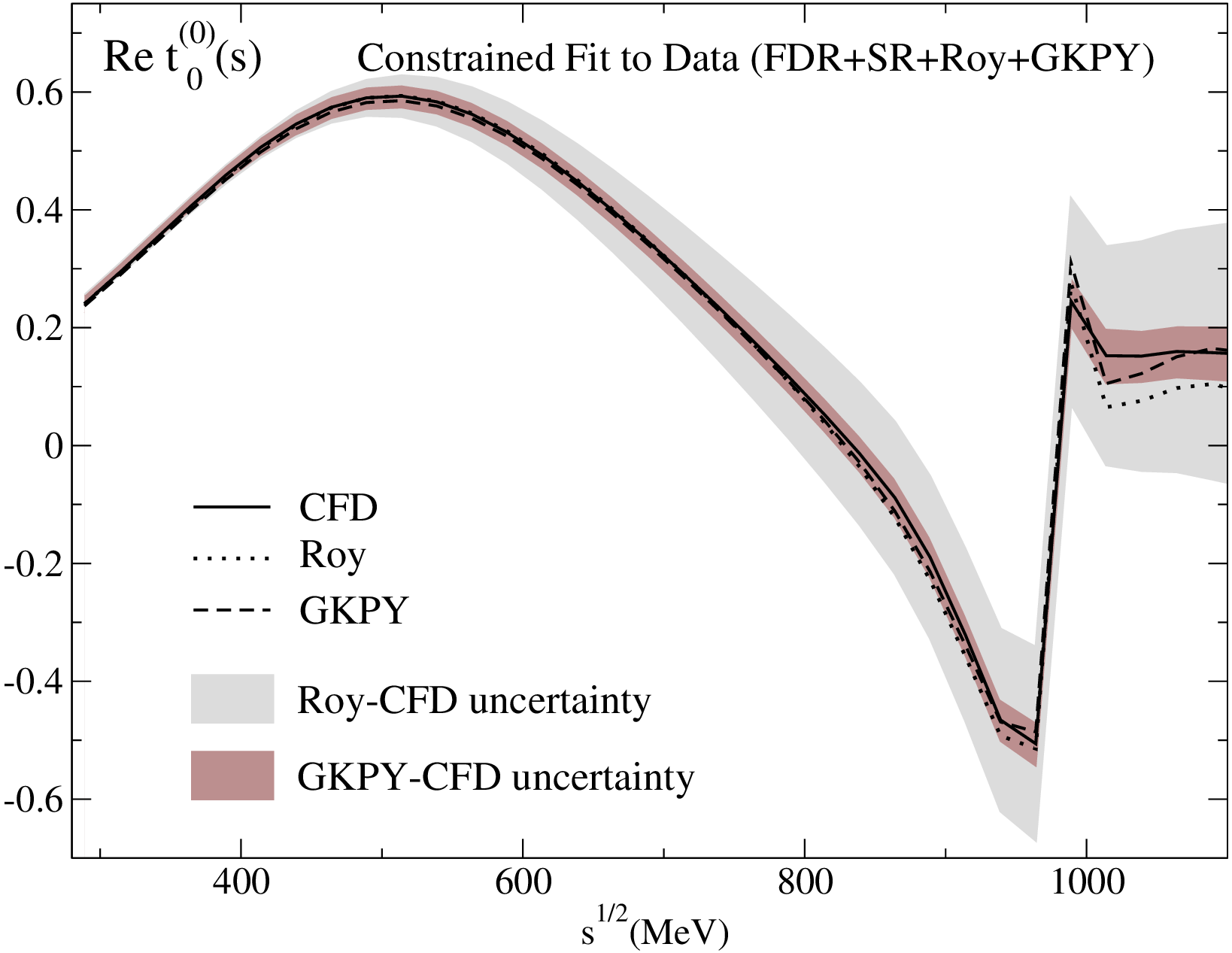}
\caption{\rm Fulfillment of S0 wave Roy and GKPY equations. 
The CFD parametrization is the input to both the Roy and GKPY eqs.,
and is in remarkable agreement with their output.
Note how the uncertainty in the Roy eq. is much larger 
than that of the GKPY eq. above roughly 500 MeV.
}
\label{RoyGKPY}
\end{figure}

The interest of this CFD parametrization is that, 
while describing the data, it satisfies within uncertainties Roy and GKPY relations 
up to their applicability range, namely 1100 MeV, 
which includes the $f_0(980)$ region. In addition, the three 
Forward Dispersion Relations are satisfied up to 1420 MeV.  
In Fig.~\ref{RoyGKPY} we show the fulfillment
of the S0 wave Roy and GKPY eqs. and how, as explained above,
the uncertainty in the Roy eq.
is much larger than for the GKPY eq. in the resonance region.
The latter will allow us now to obtain a precise determination
of the $f_0(600)$ and $f_0(980)$ poles from data alone, i.e. without using ChPT predictions.

Hence, we now feed our CFD parameterizations as input for the GKPY and Roy eqs., which provide a model independent analytic continuation to the complex plane, and
determine the position and residues of the second 
Riemann sheet poles. 
It has been shown \cite{Caprini:2005zr} that the $f_0(600)$ and $f_0(980)$
poles lie well within the domain of validity of Roy equations, given by the constraint that the $t$ values
which are integrated to obtain the partial wave representation at a given $s$ should be contained within a Lehmann-Martin ellipse. These are conditions on the analytic extension of the partial wave expansion, unrelated to
the number of subtractions in the dispersion relation, and equally apply to GKPY eqs.

Thus, in Table~\ref{tab:polesCFD} we show the $f_0(600)$, $f_0(980)$ and $\rho(770)$ poles
resulting from the use of the CFD parametrization inside Roy or GKPY eqs. 
 We 
consider that our best results are those coming from GKPY eqs. since
their uncertainties
are smaller,
although, of course, both results are compatible.

Several remarks are in order. First, statistical uncertainties are calculated using a MonteCarlo Gaussian sampling of the CFD parameters
with 7000 samples distributed within 3 standard deviations. A systematic uncertainty
due to the different charged and neutral kaon masses is relevant for the  $f_0(980)$ 
due to the existence of two $\bar KK$ thresholds separated by  roughly 8 MeV, 
which we have treated as a single $\bar KK$ threshold
at $\hat m_K=(m_{K^0}-m_{K^+})/2\simeq 992\,$MeV. In order to estimate this
systematic uncertainty, we have refitted the UFD and CFD sets to the extreme cases 
of using $m_{K^0}$ or $m_{K^+}$ instead of $\hat m_K$.
As it could be expected, the only significant variation is for the $f_0(980)$---actually, only for its half width,
which  changes by $\pm4.4\,$ MeV for GKPY eqs, and $\pm5.6\,$MeV for Roy eqs. 
The  $f_0(600)$ changes by roughly 1 MeV and the $\rho(770)$ barely notices the change---less than 0.1 MeV. The effect on residues is smaller than that of rounding the numbers. 
We have added all these uncertainties in quadrature to the statistical ones. 
Second, both the mass and width
of the $f_0(600)$ are compatible with those in ref.\cite{Caprini:2005zr} within
one standard deviation. Since we are not using ChPT and ref.\cite{Caprini:2005zr}
did not use data below 800 MeV, this is
a remarkable check of the agreement between ChPT and low energy data.
Third, the $f_0(980)$ width is no longer so narrow---as it happens in typical $\pi\pi$ scattering analyses--- and we find $\Gamma=50^{+20}_{-12} $~ MeV, 
very compatible with results from production processes. The mass overlaps
within one standard deviation with the PDG estimate.
These results show that the effect 
of the too narrow $f_0(980)$ pole 
and the use of further theoretical input like ChPT do not affect 
significantly the resulting $f_0(600)$ parameters.

In Table~\ref{tab:polesCFD} we also 
provide for each resonance its coupling to two pions, defined from its pole residue as:
\begin{equation}
g^2=-16\pi \!\!\lim_{s\rightarrow s_{pole}}\!\!(s-s_{pole})\,t_{\ell}(s)\,(2\ell+1)/(2p)^{2\ell} 
\end{equation}
where $p^2=s/4-m_\pi^2$. This residue is relevant for models of the
spectroscopic nature of these particles, particularly for the $f_0(600)$
\cite{sigmacoupling}, which are beyond the pure data analysis scope of this work.
Differences between previous values of these couplings can be seen in Tables \ref{tab:sigmapolecoupling} and \ref{tab:f0polecoupling}.

\begin{table}
\begin{tabular}{|c|c|c|}\hline
&$\sqrt{s_{\rm{pole}}}$ (MeV)&$\vert g \vert$\\\hline
\rule[-2mm]{0mm}{6mm}$f_0(600)^{\rm Roy}$ & $(445\pm25)-i(278^{+22}_{-18})$&$3.4\pm0.5$ GeV\\
\rule[-2mm]{0mm}{6mm}$f_0(980)^{\rm Roy}$ & $(1003^{+5}_{-27})-i(21^{+10}_{-8})$ &$2.5^{+0.2}_{-0.6}$ GeV\\
\rule[-2mm]{0mm}{6mm}$\rho(770)^{\rm Roy}$ & $(761^{+4}_{-3})-i(71.7^{+1.9}_{-2.3})$& $5.95^{+0.12}_{-0.08}$\\\hline
\hline
\rule[-2mm]{0mm}{6mm}$f_0(600)^{\rm GKPY}$ & $(457^{+14}_{-13})-i(279^{+11}_{-7})$& $3.59^{+0.11}_{-0.13}$ GeV\\
\rule[-2mm]{0mm}{6mm}$f_0(980)^{\rm GKPY}$ & $(996\pm7)-i(25^{+10}_{-6})$&$2.3\pm0.2$ GeV\\
\rule[-2mm]{0mm}{6mm}$\rho(770)^{\rm GKPY}$ & $(763.7^{+1.7}_{-1.5})-i(73.2^{+1.0}_{-1.1})$&$6.01^{+0.04}_{-0.07}$\\\hline
\end{tabular}\caption{Poles and residues from Roy and GKPY eqs.}
\label{tab:polesCFD}
\end{table}

In summary, using a recently developed dispersive formalism, which
is especially accurate in the resonance region,
we have been able to determine, in a model independent way, 
the $f_0(600)$, $f_0(980)$ poles and couplings from data 
with no further theoretical input.
We hope this works helps clarifying the somewhat controversial
situation regarding the parameters of these resonances.


\end{document}